\documentstyle[12pt]{article}

\begin{document}
\hbadness=10000
\hbadness=10000
\begin{titlepage}
\nopagebreak
\begin{flushright}
{\normalsize
DPSU-00-04\\
December, 2000
}
\end{flushright}
\vspace{0.7cm}
\begin{center}

{\large \bf Split Multiplets, Coupling Unification\\
and Extra Dimension}

\vspace{1.2cm}

{Yoshiharu Kawamura}

\vspace{0.9cm}
Department of Physics, Shinshu University,
Matsumoto 390-8621, Japan

\end{center}
\vspace{0.9cm}

\nopagebreak

\begin{abstract}
We study a gauge coupling unification scenario 
based on a non-supersymmetric 5-dimensional model.
Through an orbifold compactification, we obtain the 
Standard Model with split multiplets on a 4-dimensional wall, 
which is compatible with a grand unification.
\end{abstract}
\vfill
\end{titlepage}
\pagestyle{plain}
\newpage
\def\thefootnote{\fnsymbol{footnote}}
The Standard Model (SM) has been established as an effective theory
below the weak scale.
One of intriguing trials beyond the SM is to unify
gauge interactions under a simple group such as $SU(5)$.\cite{GUT}
This scenario is very attractive,\cite{GUT2} but it suffers from
several problems in the simplest version.
First problem is that gauge coupling constants do not meet 
at a high-energy scale based on the desert hypothesis.\cite{LEP}
Second problem concerns that a dangerous proton decay 
is induced by an exchange of $X$ and $Y$ gauge bosons.\cite{proton}
Last problem regards that the weak scale is not stabilized by quantum
corrections (the gauge hierarchy problem).\cite{ghp}

The introduction of supersymmetry (SUSY) solves the first \cite{unification}
and the third problems.\cite{SUSY}\footnote{
The gauge hierarchy problem is solved partially in a sence that
the hierarchy is stabilized against radiative corrections perturbatively 
due to non-renormalization theorem, although 
the origin at the tree level is not understood.}
A supersymmetric grand unified theory (SUSY GUT) is an attractive
candidate of a high-energy theory,\cite{SUSY GUT}
but the proton stability is threatened due to a contribution
from the dimension 5 operator in the minimal SUSY $SU(5)$
GUT.\cite{dim5,proton2}
Recently stronger constraints have been obtained from the analysis including
a Higgsino dressing diagram with right-handed matter fields.\cite{GN}

Recently a new possibility\cite{K2} has been proposed to solve the above
problems.
Starting from 5-dimensional (5D) SUSY $SU(5)$ model, we have obtained a
low-energy
theory with particles of the minimal supersymmetric standard model (MSSM) 
on a 4D wall through compactification upon $S^1/(Z_2 \times Z_2')$.
The proton stability is guaranteed
due to a suppression factor in the coupling
to the Kaluza-Klein modes if our 4D wall fluctuates flexibly.

In this letter, we propose another possibility to solve the first and
second problems
based on a 5D model without SUSY.
The gauge coupling unification is realized by the introduction of extra
multiplets which 
split after an orbifold compactification.\footnote{
There are several works with respect to $SU(5)$ grand unification under
an assumption that there are split multiplets.\cite{sp1,sp2,sp3}}
The splitting originates from a non-universal parity assignment on
a compact space among components in each multiplet.
The proton decay is suppressed enough with a suppression factor in the coupling
to the Kaluza-Klein excitations if our 4D wall fluctuates pliantly.
In the following, we will derive a model discussed in Ref.\cite{sp3}
as an example of a low-energy theory from 5D $SU(5)$ model 
through compactification upon $S^1/(Z_2 \times Z_2')$.

The space-time is assumed to be factorized into a product of 
4D Minkowski space-time $M^4$ and the orbifold $S^1/(Z_2 \times
Z_2')$,\footnote{
Recently, Barbieri, Hall and Nomura have
constructed a constrained standard model 
upon a compactification of 5D SUSY model on 
the orbifold $S^1/(Z_2 \times Z_2')$.\cite{BHN}
They have used a $Z_2 \times Z_2'$ parity to reduce SUSY.
There had been several works on the model building 
through a reduction of SUSY \cite{AMQ,HW,MP,PQ} or a gauge symmetry \cite{K} 
by the use of a $Z_2$ parity.
Attempts to construct GUT have been made through the
dimensional reduction over coset space.\cite{coset}}
whose coordinates are denoted
by $x^\mu$ ($\mu = 0,1,2,3$) and $y(=x^5)$, respectively.
The 5D notation $x^M$ ($M = 0,1,2,3,5$) is also used.
The orbifold is regarded as 
an interval with a distance of $\pi R/2$.
There are two 4D walls placed at
fixed points $y=-\pi R/2$ and $y=0$ 
(or $y'=0$ and $y'=\pi R/2$) on $S^1/(Z_2 \times Z_2')$
where $y' \equiv y + \pi R/2$.

We assume that the 5D gauge boson $A_M(x^\mu,y)$ and four kinds of scalar
fields $\Phi_{{\bf R}}(x^\mu,y)$ (${\bf R} =$${\bf 5}$, 
$\overline{\bf 5}$, ${\bf 10}$, $\overline{\bf 10}$)
exist in the bulk $M^4 \times S^1/(Z_2 \times Z_2')$.
The fields $A_M$ and $\Phi_{\bf R}$ form an adjoint representation 
${\bf 24}$ and a representation ${\bf R}$ of $SU(5)$, respectively.
We assume that our visible world is one of 4D walls
(We choose the wall fixed at $y=0$ as the visible one and call it wall I)
and that three families of quarks and leptons,
$3\{\psi_{\overline{\bf 5}} + \psi_{\bf 10}\}$, are located on wall I.
(Here and hereafter we suppress the family index.)
The representations of $\psi_{\overline{\bf 5}}$ and $\psi_{\bf10}$ 
are $\overline{\bf 5}$ and ${\bf 10}$ of $SU(5)$, respectively.
Note that matter fields contain no excited states along the
$S^1/(Z_2 \times Z_2')$ direction.

The gauge invariant action is given by
\begin{eqnarray}
 S &=& \int d^5x \left( - {1 \over 2} {\rm tr} F_{MN}^2 + 
\sum_{\bf R} |D_M \Phi_{\bf R}|^2 - V(\Phi_{\bf R}) \right) \nonumber\\
&~&  + \int d^4x \sum_{3 {\rm families}} ( i \overline{\psi}_{\bf 10} 
\gamma^\mu D_\mu {\psi}_{\bf 10}
    + i \overline{\psi}_{\overline{\bf 5}} \gamma^\mu D_\mu
{\psi}_{\overline{\bf 5}} 
\nonumber\\
&~& + f_{U(5)} \Phi_{\bf 5} \psi_{\bf10} \psi_{\bf10} 
 + f_{D(5)} \Phi_{\overline{\bf 5}} \psi_{\bf10} \psi_{\overline{\bf 5}} 
 + f_{Q(5)} \Phi_{\bf 10} \psi_{\overline{\bf 5}} \psi_{\overline{\bf 5}} +
\mbox{h.c.} )  , 
\end{eqnarray}
where $D_M \equiv \partial_M - i g_{(5)} A_M(x^\mu, y)$,
$g_{(5)}$ is a 5D gauge coupling constant, and $f_{U(5)}$, $f_{D(5)}$ and
$f_{Q(5)}$
are 5D Yukawa coupling matrices.
In 4D action, the bulk fields $A_M$ and $\Phi_{\bf R}$ are replaced
by fields including the Nambu-Goldstone boson $\phi(x^\mu)$ at wall I such that 
$A_M(x^\mu, \phi(x^\mu))$ and $\Phi_{\bf R}(x^\mu, \phi(x^\mu))$, respectively.
The Lagrangian is invariant under
the $Z_2 \times Z_2'$ transformation
\begin{eqnarray}
 &~& A_{\mu}(x^\mu, y) \to A_{\mu}(x^\mu, -y) = 
     P A_{\mu}(x^\mu, y) P^{-1} , \nonumber \\
 &~& A_{5}(x^\mu, y) \to A_{5}(x^\mu, -y) = 
     - P A_{5}(x^\mu, y) P^{-1} , \nonumber \\
 &~& \Phi_{\bf R}(x^\mu, y) \to \Phi_{\bf R}(x^\mu, -y) = P \Phi_{\bf
R}(x^\mu, y)  ,
  ~~~~~~~~~~ ({\bf R}={\bf 5},\overline{\bf 5})\nonumber \\ 
 &~& \Phi_{\bf R}(x^\mu, y) \to \Phi_{\bf R}(x^\mu, -y) = P \Phi_{\bf
R}(x^\mu, y) P^{-1} , 
  ~~~~~ ({\bf R}={\bf 10},\overline{\bf 10})
\label{P-tr2} \\
 &~& A_{\mu}(x^\mu, y') \to A_{\mu}(x^\mu, -y') = 
      P' A_{\mu}(x^\mu, y') P'^{-1} , \nonumber \\
 &~& A_{5}(x^\mu, y') \to A_{5}(x^\mu, -y') = 
      - P' A_{5}(x^\mu, y') P'^{-1} , \nonumber \\
 &~& \Phi_{\bf R}(x^\mu, y') \to \Phi_{\bf R}(x^\mu, -y') = P' \Phi_{\bf
R}(x^\mu, y')  ,
  ~~~~~~~~ ({\bf R}={\bf 5},\overline{\bf 5})\nonumber \\ 
 &~& \Phi_{\bf R}(x^\mu, y') \to \Phi_{\bf R}(x^\mu, -y') = P' \Phi_{\bf
R}(x^\mu, y') P'^{-1}  ,
  ~~ ({\bf R}={\bf 10},\overline{\bf 10})  
\label{P'-tr2}
\end{eqnarray}
where $P$ and $P'$ are $5 \times 5$ matrices which satisfy 
$P^2 =  P'^2 = I$ ($I$ is the unit matrix).
Here $A_M$ and $\Phi_{{\bf 10}(\overline{\bf 10})}$ are expressed by $5
\times 5$ 
symmetric and anti-symmetric matrices, respectively.
The intrinsic $Z_2 \times Z_2'$ parity of each component is given by an eigenvalue of $P$
and $P'$.

When we take $P={\rm diag}(1,1,1,1,1)$ and $P'={\rm
diag}(-1,-1,-1,1,1)$,\footnote{
The exchange of $P$ for $P'$ is equivalent to 
the exchange of two walls.}
the $SU(5)$ gauge symmetry is reduced to that of
the SM, $G_{\rm SM} \equiv SU(3) \times SU(2)
\times U(1)$, in 4D theory.\footnote{
Our symmetry reduction mechanism is 
different from the Hosotani mechanism.\cite{H}
In fact, the Hosotani mechanism does not work in our case,
because $A_{5}^{a}(x^\mu,y)$ has odd parity, as given in (\ref{P-tr2}),
and its VEV should vanish.}
This is because the boundary conditions on $S^1/(Z_2 \times Z_2')$ 
given in (\ref{P'-tr2}) do not respect $SU(5)$ symmetry,
as we see from the relations for the gauge generators $T^A$ 
$(A = 1, 2,...,24)$,
\begin{eqnarray}
  P' T^a P'^{-1} = T^a , ~~
  P' T^{\hat{a}} P'^{-1} = -T^{\hat{a}}  .
\end{eqnarray}
The $T^a$s are gauge generators of $G_{\rm SM}$
and the $T^{\hat{a}}$s are other gauge generators.
The parity assignment and mass spectrum after compactification are given in
Table 1.
\begin{table}[b]
\caption{Parity and Mass spectrum.}
\begin{center}
\begin{tabular}{l|l|l|l}
\hline\hline
4D fields & Quantum numbers & $Z_2 \times Z_2'$ parity & Mass \\
\hline
$A_{\mu}^{a(2n)} $ & $({\bf 8}, {\bf 1}) + ({\bf 1}, {\bf 3})
 + ({\bf 1}, {\bf 1})$ & $(+, +)$ & $\displaystyle{{2n \over R}}$ \\
$A_{\mu}^{\hat{a}(2n+1)}$ &  $({\bf 3}, {\bf 2}) + (\overline{\bf 3}, {\bf 2})$
& $(+, -)$ & $\displaystyle{{2n+1 \over R}}$ \\
\hline
$A_{5}^{a(2n+2)}$ & $({\bf 8}, {\bf 1}) + ({\bf 1}, {\bf 3})
 + ({\bf 1}, {\bf 1})$ & $(-, -)$ & $\displaystyle{{2n+2 \over R}}$ \\
$A_{5}^{\hat{a}(2n+1)}$ &  
$({\bf 3}, {\bf 2}) + (\overline{\bf 3}, {\bf 2})$
& $(-, +)$ & $\displaystyle{{2n+1 \over R}}$ \\
\hline
$\phi_C^{(2n+1)}$ & $({\bf 3}, {\bf 1})$ & $(+, -)$ & $\displaystyle{{2n+1
\over R}}$ \\
$\phi_W^{(2n)}$ & $({\bf 1}, {\bf 2})$ & $(+, +)$ & $\displaystyle{{2n
\over R}}$ \\
\hline
$\overline{\phi}_{C}^{(2n+1)}$ & $(\overline{{\bf 3}}, {\bf 1})$ & $(+, -)$ 
& $\displaystyle{{2n+1 \over R}}$ \\
$\overline{\phi}_{W}^{(2n)}$ & $({\bf 1}, {\bf 2})$ & $(+, +)$ &
$\displaystyle{{2n \over R}}$ \\
\hline
$Q^{(2n)}$ & $({\bf 3}, {\bf 2})$ & $(+, +)$ & $\displaystyle{{2n \over R}}$ \\
${\overline{U}}^{(2n+1)}$ & $(\overline{{\bf 3}}, {\bf 1})$ & $(+, -)$ 
& $\displaystyle{{2n+1 \over R}}$ \\
$\overline{E}^{(2n+1)}$ & $({\bf 1}, {\bf 1})$ & $(+, -)$ &
$\displaystyle{{2n+1 \over R}}$ \\
\hline
$\overline{Q}^{(2n)}$ & $(\overline{{\bf 3}}, {\bf 2})$ & $(+, +)$ &
$\displaystyle{{2n \over R}}$ \\
${U}^{(2n+1)}$ & $({\bf 3}, {\bf 1})$ & $(+, -)$ 
& $\displaystyle{{2n+1 \over R}}$ \\
$E^{(2n+1)}$ & $({\bf 1}, {\bf 1})$ & $(+, -)$ & $\displaystyle{{2n+1 \over
R}}$ \\
\hline
\end{tabular}
\end{center}
\end{table}
The scalar fields $\Phi_{\bf R}$ are broken up into several pieces such that
\begin{eqnarray}
&~& \Phi_{\bf 5} = \phi_C + \phi_W , ~~~~ \Phi_{\overline{\bf 5}} 
= \overline{\phi}_{C} + \overline{\phi}_{W} , 
\nonumber \\
&~& \Phi_{\bf 10} = Q + \overline{U} + \overline{E} , ~~~~ 
\Phi_{\overline{\bf 10}} = \overline{Q} + U + E .
\end{eqnarray}
In the second column, we give $SU(3) \times SU(2)$ quantum numbers
of 4D fields.
In the third column, $(\pm, \pm)$ and $(\pm, \mp)$ denote
eigenvalues $(\pm 1, \pm1)$ and $(\pm 1, \mp1)$ 
of $Z_2 \times Z_2'$ parity, respectively.
The fields $\phi_{\pm\pm}(x^\mu,y)$ and $\phi_{\pm\mp}(x^\mu,y)$, whose
values of  
intrinsic parity are $(\pm 1, \pm1)$ and $(\pm 1, \mp1)$, 
are Fourier expanded as
\begin{eqnarray}
  \phi_{++} (x^\mu, y) &=& 
       \sqrt{2 \over {\pi R}} 
      \sum_{n=0}^{\infty} \phi^{(2n)}_{++}(x^\mu) \cos{2ny \over R} ,
\label{phi++exp}\\
  \phi_{+-} (x^\mu, y) &=& 
       \sqrt{2 \over {\pi R}} 
      \sum_{n=0}^{\infty} \phi^{(2n+1)}_{+-}(x^\mu) \cos{(2n+1)y \over R} ,
\label{phi+-exp}\\
  \phi_{-+} (x^\mu, y) &=& 
       \sqrt{2 \over {\pi R}}
      \sum_{n=0}^{\infty} \phi^{(2n+1)}_{-+}(x^\mu) \sin{(2n+1)y \over R}  ,
\label{phi-+exp}\\
  \phi_{--} (x^\mu, y) &=& 
       \sqrt{2 \over {\pi R}}
      \sum_{n=0}^{\infty} \phi^{(2n+2)}_{--}(x^\mu) \sin{(2n+2)y \over R}  ,
\label{phi--exp}
\end{eqnarray}
where $n$ is zero or a positive integer, and each field
$\phi^{(2n)}_{++}(x^\mu)$,
$\phi^{(2n+1)}_{\pm\mp}(x^\mu)$  and $\phi^{(2n+2)}_{--}(x^\mu)$
acquire a mass $\displaystyle{{2n \over R}}$, 
$\displaystyle{{2n+1 \over R}}$ and 
$\displaystyle{{2n+2 \over R}}$ upon compactification. 
Note that 4D massless fields appear only from components with even parity
$(+1, +1)$.
The contribution from the potential $V(\Phi_{\bf R})$
is not considered in the fourth column.
In the low-energy spectrum, there are a pair of lepto-quark bosons
$Q^{(0)}$ and $\overline{Q}^{(0)}$, 
which have both color and weak charge.
The SM gauge bosons and the weak Higgs doublet are equivalent to
$A_{\mu}^{a(0)}$
and  $\phi_W^{(0)}$ (or $\overline{\phi}_W^{(0)}$), respectively.
The mass split of bosons is realized
by the $Z_2 \times Z'_2$ projection.

After integrating out the fifth dimension,
we obtain the 4D Lagrangian density
\begin{eqnarray}
{\cal L}^{(4)}_{\rm eff} &=&  - {1 \over 4} \sum_a {F_{\mu\nu}^{a(0)}}^2
 + |D_\mu \phi_W^{(0)}|^2 + |D_\mu \overline{\phi}_{W}^{(0)}|^2 \nonumber\\
 &~& + |D_\mu Q^{(0)}|^2 + |D_\mu \overline{Q}^{(0)}|^2
 - V(\phi_W^{(0)}, \overline{\phi}_{W}^{(0)}, Q^{(0)}, \overline{Q}^{(0)}) 
\nonumber\\
&~&  + \sum_{3 {\rm families}} 
( i \overline{\psi}_{\bf 10} \gamma^\mu D_\mu {\psi}_{\bf 10}
 + i \overline{\psi}_{\overline{\bf 5}} \gamma^\mu D_\mu
{\psi}_{\overline{\bf 5}} 
\nonumber\\
 &~& + f_U \phi_W^{(0)} q {\overline u} + f_D {\phi}_{\overline{W}}^{(0)} q
{\overline d} 
 + f_D {\phi}_{\overline{W}}^{(0)} l {\overline e}  + f_Q Q l \overline{d}
+ \mbox{h.c.}) + \cdots  ,
\label{4D-L}
\end{eqnarray}
where $D_\mu \equiv \partial_\mu - i g_{U} A_{\mu}^{(0)}$,
the dots represent terms including 
Kaluza-Klein modes, $g_U$ 
$(\equiv \sqrt{2/\pi R} g_{(5)})$ is a 4D 
gauge coupling constant, $f_U$ 
$(\equiv \sqrt{2/\pi R} f_{U(5)})$,
$f_D$ $(\equiv \sqrt{2/\pi R} f_{D(5)})$
and $f_Q$ $(\equiv \sqrt{2/\pi R} f_{Q(5)})$
are 4D Yukawa coupling matrices, and
$q$, $\overline{u}$ and $\overline{d}$ are quarks, 
$l$ and $\overline{e}$ are leptons.
With our parity assignment,
we have obtained an extention of the SM with two Higgs doublets 
$\phi_{W}^{(0)}$ and $\overline{\phi}_{W}^{(0)}$
and extra lepto-quark bosons $Q^{(0)}$ and $\overline{Q}^{(0)}$.

The theory predicts that coupling constants are unified 
around the compactification scale $M_C (\equiv 1/R)$,
as in the ordinary $SU(5)$ GUT,\cite{GUT}
\begin{eqnarray}
&~& g_3 = g_2 = g_1 = g_U , ~~ f_d = f_e = f_D~,
\end{eqnarray}
where $f_d$ and $f_e$ are Yukawa coupling matrices on
down-type quarks and electron-type leptons, respectively. 
As shown in Ref.\cite{sp3}, this type of extention of the SM can survive
with the precision measurements at LEP.\cite{LEP}

It is known that there is a significant contribution to the proton decay
due to the $X$ and $Y$ gauge bosons in the minimal $SU(5)$ GUT.\cite{proton}
In our model, we have similar diagrams as those in the minimal $SU(5)$
GUT because quark and lepton couple to the Kaluza-Klein modes
of extra vector bosons at the tree level.
However we expect that the proton stability guarantees if our 4D wall
fluctuates flexibly.
This is due to the fact 
that there is an exponential suppression factor in the coupling
to the Kaluza-Klein exicitations by the brane recoil effect.\cite{BKNY}

We have obtained the simplest extention of the SM compatible with
$SU(5)$ grand unification.
It would be possible to construct more complex models by increasing
a number of extra multiplets.
For reference, the pattern of split due to $Z_2'$ parity is
\begin{table}[b]
\caption{Split due to $Z_2'$ parity.}
\begin{center}
\begin{tabular}{r|l|r}
\hline\hline
${\bf R}$  & Quantum numbers & $Z_2'$ parity  \\
\hline
${\bf 5}$ & $({\bf 3}, {\bf 1})$ & $P_{\bf 5}$ \\
\cline{2-3}
~~ & $({\bf 1}, {\bf 2})$ & $-P_{\bf 5}$ \\
\hline
${\bf 10}$ & $({\bf 3}, {\bf 2})$ & $P_{\bf 10}$ \\
\cline{2-3}
~~ & $(\overline{\bf 3}, {\bf 1}) + ({\bf 1}, {\bf 1})$ & $-P_{\bf 10}$ \\
\hline
${\bf 15}$ & $({\bf 3}, {\bf 2})$ & $P_{\bf 15}$ \\
\cline{2-3}
~~ & $({\bf 6}, {\bf 1}) + ({\bf 1}, {\bf 3})$ & $-P_{\bf 15}$ \\
\hline
${\bf 24}$ & $({\bf 8}, {\bf 1}) + ({\bf 1}, {\bf 3}) + ({\bf 1}, {\bf 1})$
& $P_{\bf 24}$  \\
\cline{2-3}
~~  &  $({\bf 3}, {\bf 2}) + (\overline{\bf 3}, {\bf 2})$ & $-P_{\bf 24}$ \\
\hline
${\bf 45}$ & $({\bf 8}, {\bf 2}) + (\overline{\bf 3}, {\bf 2}) + ({\bf 1},
{\bf 2})$ & $P_{\bf 45}$  \\
\cline{2-3}
~~  &  $(\overline{\bf 6}, {\bf 1}) +({\bf 3}, {\bf 1}) + (\overline{\bf
3}, {\bf 1})$ & $-P_{\bf 45}$ \\
~~  &  $+ ({\bf 3}, {\bf 3})$ & ~~ \\
\hline
${\bf 75}$ & $({\bf 6}, {\bf 2}) + (\overline{\bf 6}, {\bf 2}) + ({\bf 3},
{\bf 2})$ & $P_{\bf 75}$  \\
~~ & $+ (\overline{\bf 3}, {\bf 2})$ & ~~ \\
\cline{2-3}
~~  &  $({\bf 3}, {\bf 1}) + (\overline{\bf 3}, {\bf 1}) + ({\bf 1}, {\bf
1})$ & $-P_{\bf 75}$ \\
~~  &  $+ ({\bf 8}, {\bf 3}) + ({\bf 8}, {\bf 1})$ & ~~ \\
\hline
\end{tabular}
\end{center}
\end{table}
given in Table 2 for several low dimensional representations of $SU(5)$.
In the second column, we give $SU(3) \times SU(2)$ quantum numbers
of split multiplets.
In the third column, $P_{\bf R}$ is an 
eigenvalue of $Z_2'$ parity, i.e., 
$P_{\bf R} = 1$ or $-1$.
The table includes components which can induce a rapid nucleon decay
when they couple to quarks and leptons.

Our grand unification scenario is phenomenologically interesting
because it suggests the existence of extra split multiplets
at the weak scale.
However there is a problem of
how to break the electro-weak symmetry naturally
and how to stabilize the weak scale, that is,
our model suffers from naturalness problem.\cite{naturalness}
There would be an alternative that the extra space has a large
radius.\cite{largeR}
In this case, the low-energy gauge coupling unification is expected to be
realized 
by a power-law correction.\cite{power}

\end{document}